\documentclass[twocolumn]{article}

\usepackage[T1]{fontenc}

\usepackage[]{biblatex} 
\ifpdf \usepackage[pdftex]{graphicx} \pdfcompresslevel=9
\else \usepackage[dvips]{graphicx} \fi

\usepackage{amsmath}

\usepackage{xspace}
\newcommand{\frameworkName}{\texttt{PEMesh}\xspace}

\title{A Graphical Framework to Study the Correlation between Geometric Design and Simulation}

\author
{\parbox{\textwidth}
{\centering 
D. Cabiddu,
G. Patan\'e,
and M. Spagnuolo}
        \\
{\parbox{\textwidth}{\centering CNR-IMATI, Genova, Italy \\
      }
}
}

\begin{document}

\date{}
\maketitle
\begin{abstract}
 Partial differential equations can be solved on general polygonal and polyhedral meshes, through \emph{Polytopal Element Methods} (PEMs). Unfortunately, the relation between geometry and analysis is still unknown and subject to ongoing research to identify weaker shape-regularity criteria under which PEMs can reliably work. We propose a graphical framework to support the analysis of the relation between the geometric properties of polygonal meshes and the numerical performances of PEM solvers. Our framework, namely \frameworkName, allows the design of polygonal meshes that increasingly stress some geometric properties, by exploiting any external PEM solver, and supports the study of the correlation between the performances of such a solver and the geometric properties of the input mesh. Furthermore, it is highly modular, customisable, easy to use, and provides the possibility to export analysis results both as numerical values and graphical plots. The framework has a potential practical impact on ongoing and future research activities related to PEM methods, polygonal mesh generation and processing.
   \\



\end{abstract}  
\section{Introduction\label{sec:INTRODUCTION}}
Over the last fifty years, computer-based simulation has dramatically increased its impact on research, design, and production, and is now an indispensable tool for development and innovation in science and technology. In particular, Partial Differential Equations (PDEs) offer a broad and flexible framework for modelling and analyzing several phenomena arising in fields as diverse as physics, engineering, biology, and medicine. Computer-based simulation of PDEs also relies on a suitable description of geometrical entities, such as the computational domain and its properties. However, the representation of geometric entities has been studied mainly in the field of geometric modelling, and often the requirements of shape design are different from those of numerical simulation.

In this context, \emph{Polytopal Element Methods} (PEMs) allow solving differential equations on general polygonal and polyhedral meshes, thus offering great freedom in the definition of mesh generation algorithms. Similarly to \emph{Finite Elements Methods} (FEMs), the performance of PEMs (i.e., accuracy, stability, and effectiveness of preconditioning) depends on the quality of the underlying mesh. Differently from FEMs, where the relation between the geometric properties of the mesh and the performances of the solver is well known~\cite{shewchuk2002good, ciarlet2002, BRANDTS20082227}, the definition of the quality of polytopal elements is still subject to ongoing research~\cite{shapereg,mu2015shape,CHANGE}. 

The proposed graphical framework is intended to support the analysis of the relation between the geometric properties of the mesh and the numerical performances of the solver, in terms of basis degree, conditioning number of the stiffness matrix, etc. To this end, our work covers several aspects, such as the design and generation of meshes to increasingly stress geometric properties, the study of the performances of PEM solvers, and the correlation between such performances and the main geometric properties of the input meshes. Each step is performed by exploiting existing tools mainly coming from two related but independent research areas: geometric design and numerical methods for PEMs. Indeed, these tools rely on different representations of the same domain, and researchers are often required to be skilled programmers and expert tool users to allow such tools to be part of the same experimental pipeline. 

\begin{figure*}[h]
    \centering
\includegraphics[width=.92\textwidth]{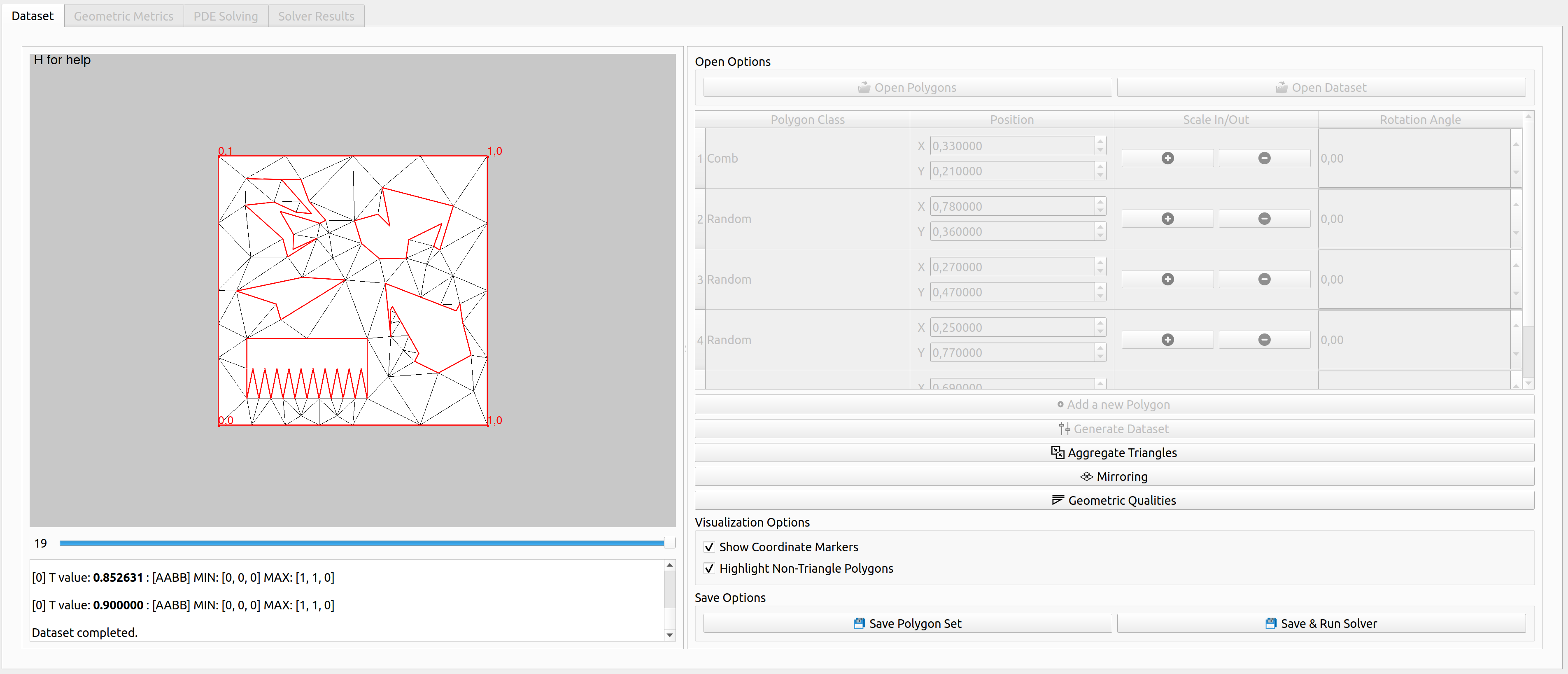}
\caption{Main window. 
On the left, a canvas allows visualizing the input mesh(es) and the bottom text box reports some logs. On the right, an advanced panel allows setting multiple parameters to both guide the dataset generation according to desired geometric features and start the geometric analysis. Also, a button is provided to quickly start the PEM solver and run the simulation.
\label{fig:app}}
\end{figure*}

We introduce \frameworkName (Fig.~\ref{fig:app}), as 
an open-source software tool designed to help researchers to perform experiments on the analysis and design of polytopal meshes for PEM solvers. 
It is an advanced graphical tool that seamlessly integrates geometric design pipelines and PEM simulations. Specifically, it supports the design and generation of complex input polygonal meshes by stressing geometric properties while providing the possibility to solve PEMs on the generated meshes. Furthermore, \frameworkName
allows the user to correlate one or more geometric properties of the input polytopal mesh with the performances of PEM solvers, and to visualize the results through customisable and interactive plots. 

The proposed tool is highly modular and customisable. It allows researchers to simulate any PEM solver, by simply calling the PEM solver executable from an internal command line and providing possibly additional input parameters other than the geometric data set. To the best of our knowledge, 
our proposal is the first graphical tool to generate complex discrete polytopal meshes and to support a study of the correlation between their geometric properties and numerical PEMs solvers. Indeed, it has a potential practical impact on research activities on this subject. 

The paper is organised as follows. We briefly review previous work on PEMs and existing tools both to perform PEM simulations and to design geometric data sets (Sect.~\ref{sec:related}). Then, we describe the structure of 
our framework and its capabilities (Sect.~\ref{sec:overview}), with technical details about its implementation (Sect.~\ref{sec:implementation}). Finally, we discuss some directions for future research (Sect.~\ref{sec:conclusions}).

\section{Background and related work\label{sec:related}}
We briefly review previous work on numerical FEM solvers and libraries (Sect.~\ref{sec:NUMERICAL-LIBRARIES}) and meshing tools (Sect.~\ref{sec:MESHING-TOOLS}).

\subsection{Numerical FEM solvers and libraries\label{sec:NUMERICAL-LIBRARIES}}
Main PEMs include Mimetic Finite Differences~\cite{book_mimetic,brezzi_mimetic}, Discontinuous Galerkin-Finite Element Method (DG-FEM)~\cite{Rev_DG,Cangiani_hpDGVEM}, Hybridisable and Hybrid High-Order Methods~\cite{Cockburn_LDG,Ern_LDG}, Weak Galerkin Method~\cite{Weak_FEM}, BEM-based FEM~\cite{BEM_FEM}, Poly-Spline FEM~\cite{schneider2019poly}, and Polygonal FEM~\cite{pol_FEM}. Main existing tools for the numerical solution of PDEs include (i) \emph{VEMLab}~\cite{vemlab}, which is an open-source MATLAB library for the virtual element method and (ii) \emph{Veamy}, which is a free and open source C++ library that implements the virtual element method (C++ version of~\cite{vemlab}). The current release of this library allows the solution of 2D linear elasto-static problems and the 2D Poisson problem~\cite{veamy}. Other libraries are (iii) the 50-lines MATLAB implementation of the lowest order virtual element method for the two-dimensional Poisson problem on general polygonal meshes~\cite{vem-50-lines}, and (iv) the MATLAB implementation of the lowest order Virtual Element Method (VEM)~\cite{mascotto2018}.

As a matter of example, we demonstrate how a PEM solver can be integrated into 
our framework. Our use case exploits the \emph{Virtual Element Method} (VEM)~\cite{basicVEM}, which can be considered as an extension to FEM for handling general polytopal meshes.

\subsection{Mesh generation tools\label{sec:MESHING-TOOLS}}
Nowadays, meshes are commonplace in several applications ranging from engineering to bio-medicine and geology. Depending on the application field, automatic mesh generation may be a difficult task, due to specific geometric requirements to be satisfied. 

Concerning simulation with FEMs, the principle behind meshing algorithms in commercial FEM solvers is described in~\cite{Okereke2018}, and an open-source tool is provided. Free-FEM~\cite{freefem} is a popular 2D and 3D partial differential equations (PDE) solver used by thousands of researchers across the world, including its mesh generation module. Although it provides plenty of functionalities, it is based on its language and it has no graphical interface. Similarly, the MATLAB\textsuperscript{\textregistered} suite provides its own FEM mesh generator~\cite{matlab-mesh-generator}. Both solutions focus on FEM requirements and enable the possibility to generate triangle meshes, but they do not allow the generation of generic polygon meshes.

Concerning PEM methods,
available Voronoi-based meshing tools (e.g.~\cite{du1999centroidal,polymesher}) are not suited for our study, because they produce convex elements that are not challenging enough to stress PEM solvers. Recent works \cite{huang2021large,sorgente2022role} propose ad-hoc polyhedral mesh datasets, specifically designed for their experimental phase. The dataset generation procedures are well described, but they are not easily modifiable to fit analyses other than the proposed approach. To the best of our knowledge, the benchmark proposed in~\cite{CHANGE} is the only one providing a polygonal mesh generation approach specifically designed for PDE solvers. Unfortunately, the proposed approach is not easily customisable and allows the generation of polygon meshes having a single non-triangle element.

Our framework provides an advanced mesh generation module which enables the creation of generic polygonal meshes specifically designed for PDE solvers.

\section{Proposed framework\label{sec:overview}}
Our framework 
is aimed at evaluating the dependence of the performances of a PEM solver on the geometrical properties of the input polygonal mesh, which is either generated by using the framework
itself or provided as an external resource. Mainly, the framework 
is composed mainly of four modules, each of which is provided as a specialized window.\\

\emph{Polygon mesh generation~$\&$ loading} allows the user to load one or more existing meshes or generate a new one from scratch by exploiting a set of provided polytopal elements or providing an external one. The generation of new meshes is highly customisable, and the user is allowed to play with a large set of options and parameters (Sect.~\ref{sec:mesh_generation}).\\

\emph{Geometric analysis} allows the user to perform a deep analysis of the geometric properties of the input polygonal meshes and to correlate each of them with the others. Results of such an analysis are shown through advanced plots (Sect.~\ref{sec:geometry_analysis}).\\

\emph{PEM solver} allows the user to run a PEM solver and analyse its performances on input polygonal meshes. Any PEM solver may be exploited, as long as it can be run from the command line and provides its output according to a specific textual format. Both the solution and the ground-through of the PEM are shown directly on the meshes, while the performances of the solver are visualized through linear plots  (Sect.~\ref{sec:pem_solvers}).\\

\emph{Correlation visualization} supports the analysis of the correlation between the geometric properties of the polygonal meshes and the numerical performances of the selected PEM solver. Results are made available in the form of customizable scatter plots (Sect.~\ref{sec:correlation}).\\

The framework 
provides the possibility to show the results of each analysis step on the display, customize visualization aspects of plots (i.e., colour, font sizes, etc.) and interactively analyze them by clicking on visualised points and lines to show data values in the selected point. Furthermore, such results can be saved on disk as images and as textual files, to be possibly re-used by other applications.

\subsection{Polygonal mesh generation~$\&$ loading\label{sec:mesh_generation}}
This module 
is started as soon as the application is run (Fig. ~\ref{fig:app}). It provides the possibility to load an existing polygonal mesh or generate a new one from scratch. 
The mesh generation method takes a cue from the approach described in~\cite{CHANGE}, where the domain is supposed to be a squared canvas, and the area of the domain which is not covered by a polygon is filled with triangles using~\cite{shewchuk1996triangle}. 

\begin{figure}[hht]
\centering
\includegraphics[width=.23\textwidth]{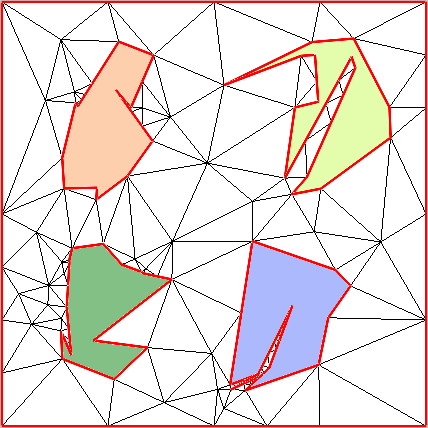}
\includegraphics[width=.23\textwidth]{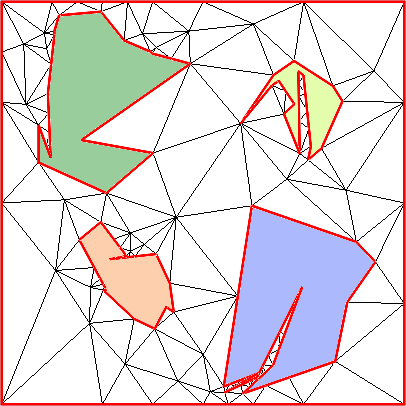}
\caption{Meshes generated by selecting the same set of polygons, but editing each element differently.\label{fig:multi-poly-rand}}
\end{figure}

Differently from~\cite{CHANGE}, our framework 
supports the generation of meshes with more than one non-triangle polygon, whose position, scale and rotation are chosen by the user before applying the triangulation of the external domain (Fig.~\ref{fig:multi-poly-rand}). Furthermore, triangulation parameters are set according to the user needs (e.g., fixing the area of the triangles or the minimum angle) and the mirroring approach proposed in~\cite{CHANGE} can be eventually applied after the triangulation.

\begin{figure}[t]
\centering
\includegraphics[width=.2\textwidth]{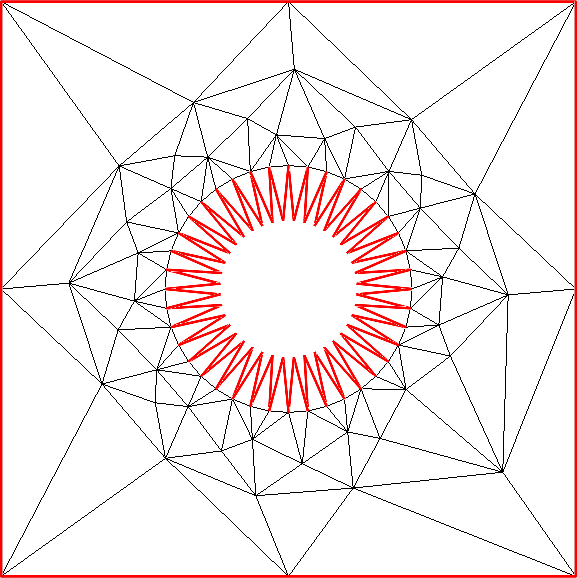}
\includegraphics[width=.2\textwidth]{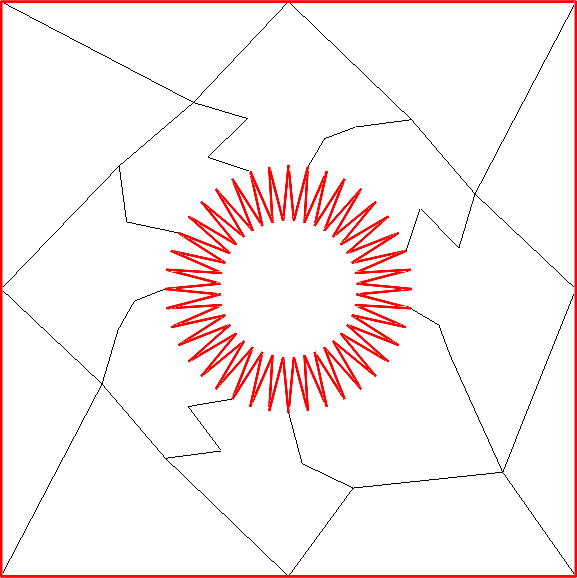}
\caption{Mesh with aggregated triangles.\label{fig:aggregated}}
\end{figure}

An additional feature is the aggregation of the generated triangles to create generic polygons (Fig.~\ref{fig:aggregated}). This feature allows the generation of generic polygonal meshes where the number of triangles is reduced almost to zero and some geometric properties are stressed all over the discretised domain. The aggregation criterion guarantees that the diameter of the polygons generated by aggregation is at most equal to the diameter of the smallest user-selected polygon. 
\subsection{Geometric analysis\label{sec:geometry_analysis}}
\begin{figure}[ht]
\centering
\begin{tabular}{cc}
\includegraphics[width=.42\columnwidth]{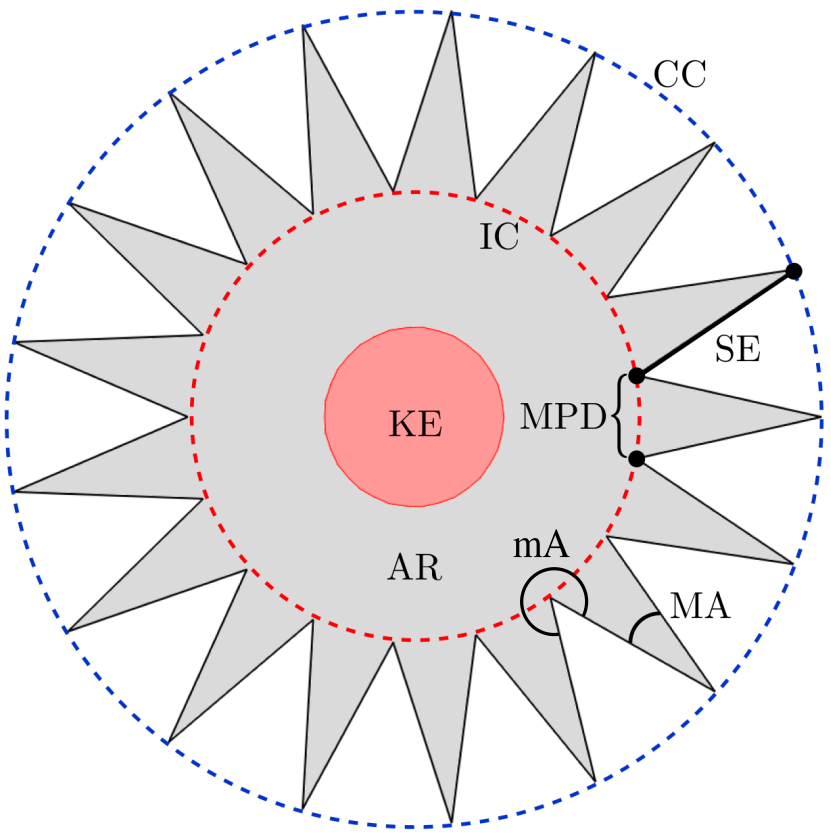}
&
\includegraphics[width=.5\columnwidth]{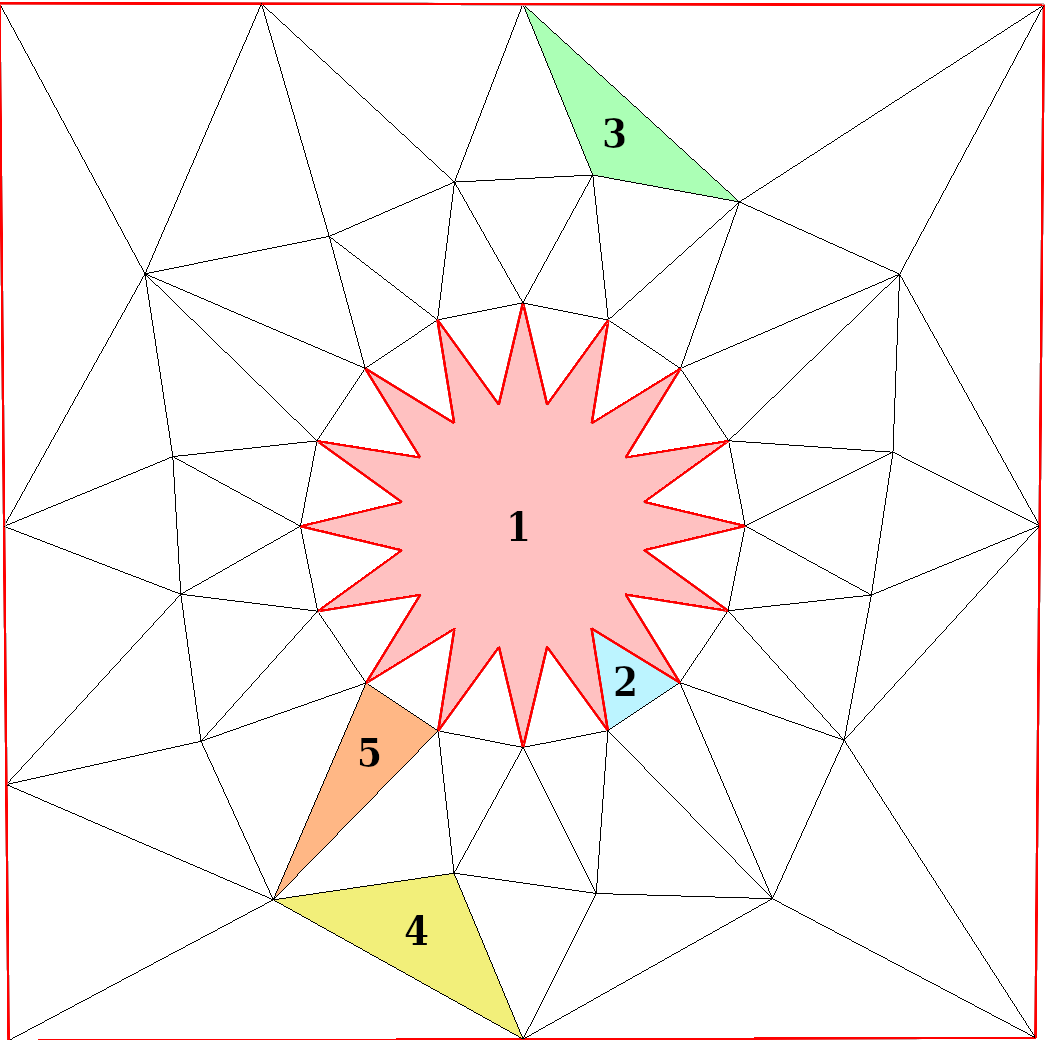}\\
(a) &(b)
\end{tabular}
\caption{(a) Geometric metrics of a polygon. (b) Minimum geometric metrics on a polygonal mesh: $\#1$ satisfies $min(MPD)$ and $min(KAR)$; $\#2$ satisfies $min(IC), min(CC), min(AR), min(KE), min(MPD), min(SE)$; $\#3$ satisfies $min(CR)$; $\#4$ satisfies $min(KAR)$; $\#5$ satisfies $min(mA)$.  \label{fig:METRICS} 
}
\end{figure}

When one or more polygonal meshes are available, either generated from scratch or loaded from disk, our framework  
allows deep geometric analysis and provides a visual summary of geometric properties. Specifically, our approach considers a set of polygonal metrics (Table~\ref{tab:geom_metrics}, Fig.~\ref{fig:METRICS}), also considering their minimum, maximum and average values. 
\begin{table}[ht]
\centering
\caption{Proposed polygonal metrics. For scale invariant measures, the fourth column indicates whether optimal values are at the top ($\uparrow$) or bottom ($\downarrow$) of the definition range. Polygon measures: inscribed circle (IC), circumscribed circle (CC), polygon area (AR), kernel area (KE), minimum angle (MA), shortest edge length (SE), and minimum point-to-point distance (MPD).\label{tab:geom_metrics}}
{\small{\begin{tabular}{|l|l|ll|l|}
\hline
\textbf{Metric} & \textbf{Abbr.} & \multicolumn{2}{c|}{\textbf{Range}} & \textbf{Scale}\\
&&&& \textbf{inv.}\\
\hline
$\sharp$ Edges & nE &$(1,+\infty)$ &-- & Yes\\
Inscribed radius & IC &$(0,\infty)$ & -- & No\\
Circumscr. radius & CC &$(0,\infty)$ & -- & No\\
Circle ratio& CR &$[0,1]$ &$\uparrow$ & Yes\\
Area & AR &$[0,\infty)$ & -- & No\\
Kernel-area &KE &$[0,\infty)$ & -- & No\\
Kernal-area ratio & KAR &$[0,1]$ &$\uparrow$ & Yes\\
Perimeter-area ratio & PAR &$(0,\infty)$ &$\uparrow$ & Yes\\
Min. angle & MA &$(0,\pi)$ &$\uparrow$ & Yes\\
Max. angle & mA &$(0,\pi)$ &$\uparrow$ & Yes\\
Shortest edge & SE &$(0,\infty)$ & -- & No\\
Edge ratio & ER &$(0,1]$ &$\uparrow$ & Yes\\
Min p2p distance & MPD &$(0,\infty)$ & -- & No\\
Normal. point dist. & NPD &$(0,1]$ &$\uparrow$ & Yes\\
Shape regularity & SRG &$(0,1]$ &$\uparrow$ & Yes\\
\hline
\end{tabular}}}
\end{table}

Our polygonal metrics are classified into 6 main classes:
\begin{itemize}
\item \emph{edges}: number of edges (nE) of the input polygon, shortest edge (SE);
\item \emph{angles}: ratio MA/mA, with MA, mA minimum, and maximum inner angle of the polygon, respectively;
\item \emph{areas}: area (AR) of the polygon, kernel area (KE), kernel-area ratio (i.e., the ratio between the area of the kernel of the polygon and its whole area), area-perimeter ratio (APR);
\item \emph{radii}: inscribed circle radius (IC), circumscribed circle radius (CC), circle ratio (CR:=IC/CC);
\item \emph{distances}: minimum point to point distance (MPD), normalized point distance (NPD) (i.e., normalized version of MPD);
\item \emph{shape regularity} (SRG), as the ratio between the radius of the circle to the circumscribed polygon and the radius of the circle inscribed in the kernel of the element.
\end{itemize}
%
Three different visualizations of the results of the geometric analysis are available, thus enabling the possibility to either analyze each mesh as a standalone object or to consider each mesh as a part of a full data set (when more polygonal meshes are available). 

\begin{figure}[h]
    \centering
    \includegraphics[width=0.9\linewidth]{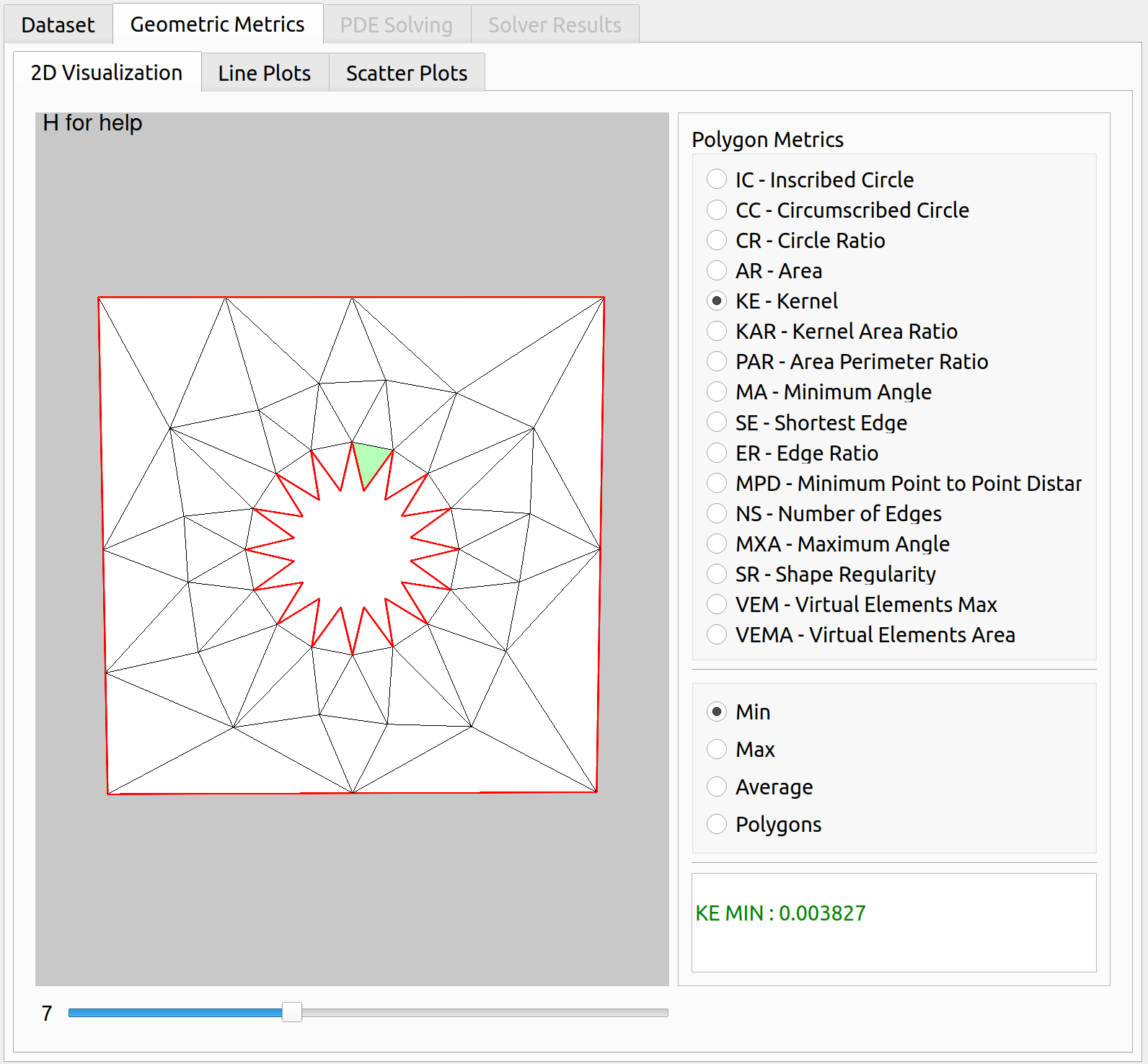}
    \caption{
    Graphical visualization of geometric polygonal metrics. Given a polygonal mesh and a selected geometric metric, the mesh element (either a triangle or a non-triangle polygon) satisfying the metric is highlighted in the canvas. Numerical values are reported in the bottom text box.
    }
    \label{fig:geometry-page}
\end{figure}

The former visualization focuses on the polygonal mesh as a standalone object (Fig.~\ref{fig:geometry-page}). On each available mesh, it visually highlights the element (either triangle or polygon) where the minimum and the maximum values are located, together with textual information about them (e.g., the numerical values). This visualization enables the detection of possible geometric degeneracies. 

\begin{figure*}[h]
    \centering
    \includegraphics[width=0.95\textwidth]{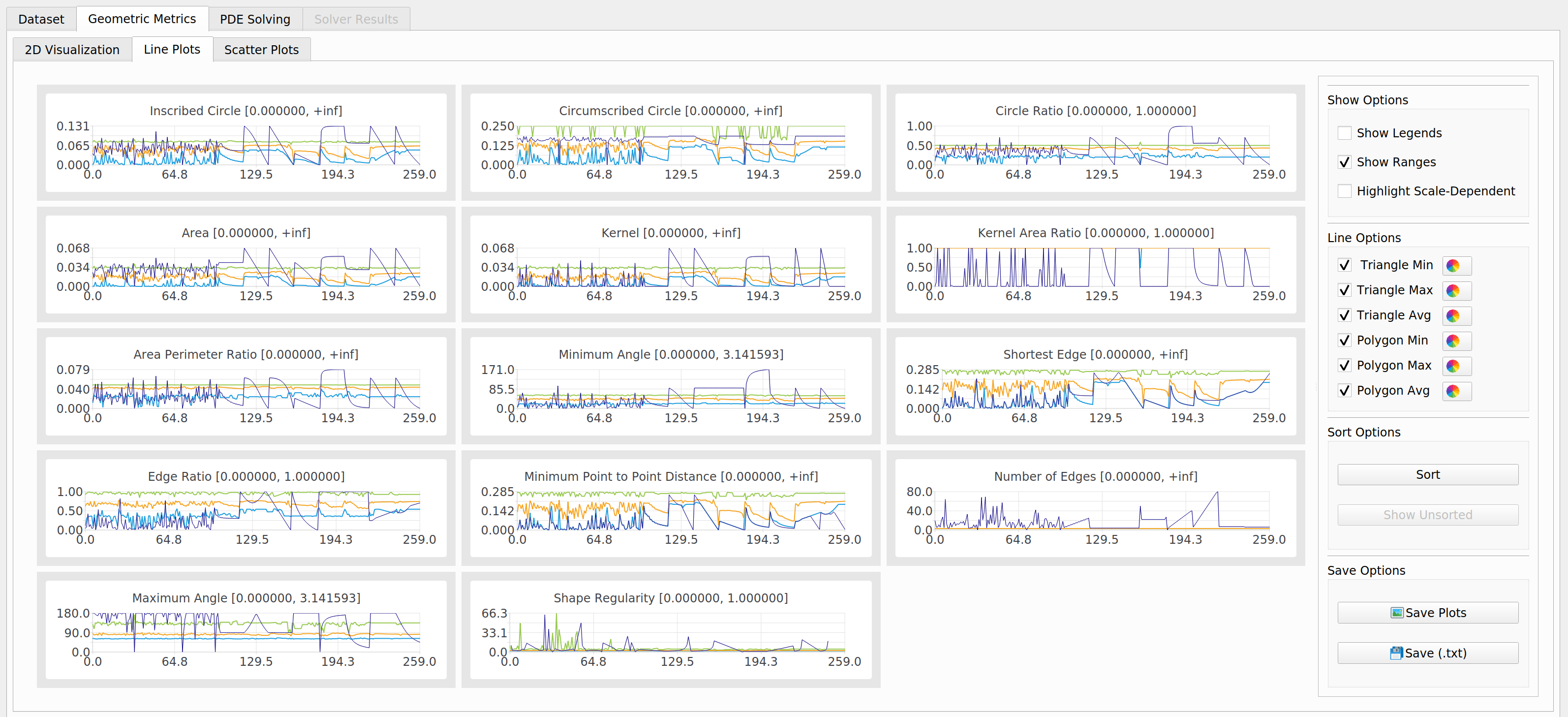}
    \caption{
    Screenshot of the window visualizing the variation of the minima and maxima of each geometric metric through line plots. By double-clicking on a single plot, the selected plot detaches from the window and enables a full-screen visualization. On the right side, a specialized panel allows graphical settings and image saving.
    }
    \label{fig:geometric-line-plots}
\end{figure*}

The other two visualization approaches consider the full data set as a single object and plot how the evaluated geometric metrics evolve in the data set, and how they correlate with each other. Specifically, a former window (Fig. ~\ref{fig:geometric-line-plots}) shows the variation of the minima and maxima of each geometric metric. Each one of these linear plots corresponds to a single geometric metric, and it is generated by setting its~$x$-axis to the indices of the meshes in the data set and its~$y$-axis to the minimum and the maximum values. The second window is specialized in correlating two user-defined geometric metrics with each other. The correlation is shown by scatter plots, each of them built by setting its axes to the two selected metrics, respectively (Sect.~\ref{sec:correlation}). In both windows, plots are interactive: a single click on a point or line in the plot shows the value corresponding to such a point. Also, both windows are highly customisable by enabling the user to set plot colours and point/line sizes. Furthermore, plots can be stored on disk as images, and numerical information can be stored as textual files.

\subsection{PEM solver\label{sec:pem_solvers}}
\frameworkName
provides the possibility to solve PEMs on a polygonal mesh of the input domain and to visualize the performances of any PEM solver. To this end, the PEM solver is not part of the tool but is rather considered an external resource that is invoked from the 
graphical interface. Without loss of generality, our framework 
assumes that the selected PEM solver takes an input mesh and returns both the solution and the ground truth (if any) of a PDE, together with statistics (e.g., approximation error, conditioning of the stiffness matrix) on the numerical solvers (Sect.~\ref{sec:pem-requirements}). 
\begin{figure*}[ht]
\centering
\includegraphics[width=.95\textwidth]{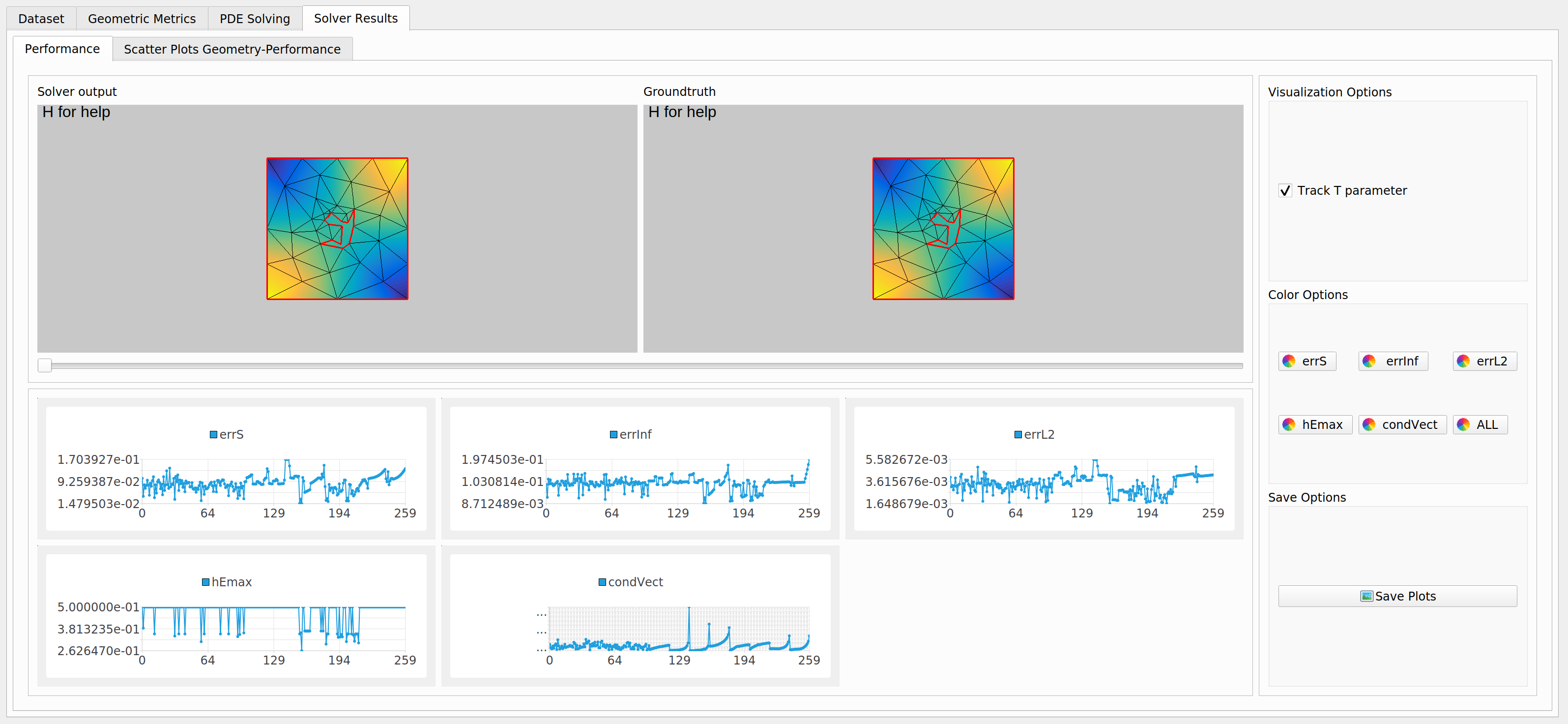}
\caption{Screenshot of the window visualizing PEM solver results. On the top, both solver output and ground truth are colour-mapped on the input polygon meshes, while on the bottom a set of linear plots show how solver performances vary in the data set.\label{fig:solver_results} 
}
\end{figure*}

When the results of the PEM solver are available, a specialized window graphically shows both the solution and the ground truth on the mesh through a colour map, while numerical and geometric metrics are represented by customisable linear plots (Fig.~\ref{fig:solver_results}). A double click on each plot detaches the plot itself from the window and enables a full-screen visualization. On the right side of the window, a set of visualization options is provided to customize both colour maps and linear plots.

\subsection{Correlation visualization\label{sec:correlation}}
\begin{figure}[h]
\centering
(a)\includegraphics[width=.33\textwidth]{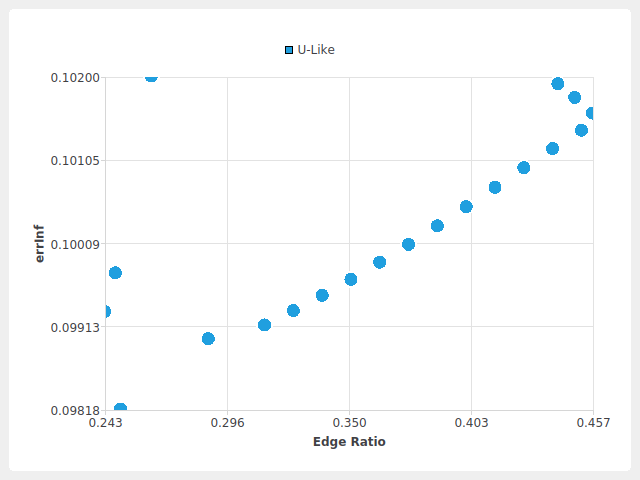}\\
(b)\includegraphics[width=.33\textwidth]{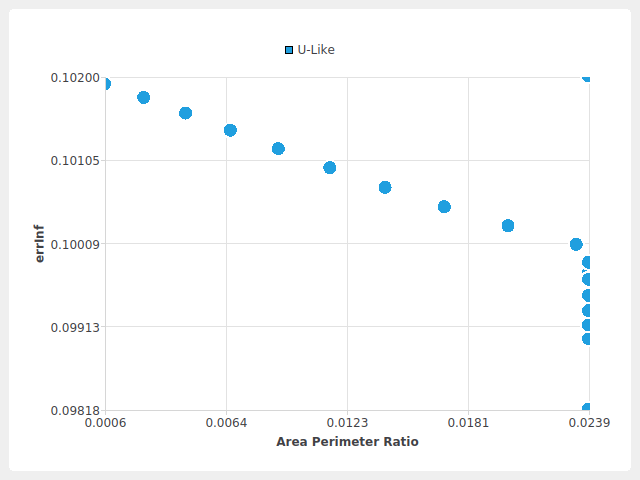}
\caption{Examples of scatter plots. (a) Positive strong correlation. (b) Negative strong correlation. \label{fig:scatterplot-pos-neg}}
\end{figure}

\frameworkName 
is intended to be a support for the investigation of possible correlations between geometric and solver performance metrics. To reach the goal, 
it provides specialized windows for the graphical visualization via scatter plots of such correlations (Fig.~\ref{fig:scatterplot-pos-neg}). A scatter plot is a plot displaying the relationship between two quantitative variables measured on the same input. The values of one variable appear on the horizontal axis, and the values of the other variable appear on the vertical axis. This representation allows us to analyse if the two variables are correlated. In the former case, it describes the relationship's direction, form, and strength. Direction can be either positive (rising) or negative (falling), while the form can be linear or curvilinear. Finally, the strength is derived from the plot's slope, indicating if the relationship is strong, moderate or weak.

\frameworkName 
provides visualization of possible correlations between two different geometric properties and between a geometric metric and a solver performance evaluation. The former is enabled when both geometric metrics and PEM performance evaluations are available.

Given two variables~$G$ and~$G'$ computed on the same set of input polygon meshes, where ~$G$ is a geometric metric and ~$G'$ may either be a geometric metric or a PEM performance evaluation, this module 
visualizes their correlation through a specialized scatter plot, with~$G$-values on the~$x$-axis and~$G'$-values on the~$y$-axis. Similarly to any other plot in the application, these scatter plots are highly customisable in terms of the point colour, size, and labels, and can be exported as images.

\section{Implementation\label{sec:implementation}}
\frameworkName
is a standalone multi-platform desktop application implemented in C++ and exploits Qt libraries for the design and implementation of the graphical user interface (Fig.~\ref{fig:app}) and Cinolib~\cite{cinolib} to generate and visualize meshes. The software we developed is open-source and publicly available. We now discuss the supported data formats (Sect.~\ref{sec:DATA-FORMAT}), the family of parametric polygons available for the generation of polygonal meshes (Sect.~\ref{sec:DATA-FORMAT}), and a description (Sect.~\ref{sec:pem-requirements}) of the PEM solver used for the experiments on the interplay between geometry and analysis.

\subsection{Supported data formats\label{sec:DATA-FORMAT}}
As mentioned, \frameworkName 
provides the possibility to either load an existing data set or generate a new one from scratch. In both cases, it 
supports the most widely used file formats for the exchange of polygonal meshes, namely OBJ, OFF and STL. Furthermore, an additional output format is provided and produces \textit{.node} and \textit{.ele} files encoding vertices and polygons respectively. This latter mesh format is provided to support a large amount of PEM solvers requiring this kind of input.

\subsection{Input polygons\label{sec:INPUT-POLYGON}}
\begin{figure}[ht]
\centering
\begin{tabular}{cc}
(a)~$t=0$ &(b)~$t=1$\\
\includegraphics[width=.22\textwidth]{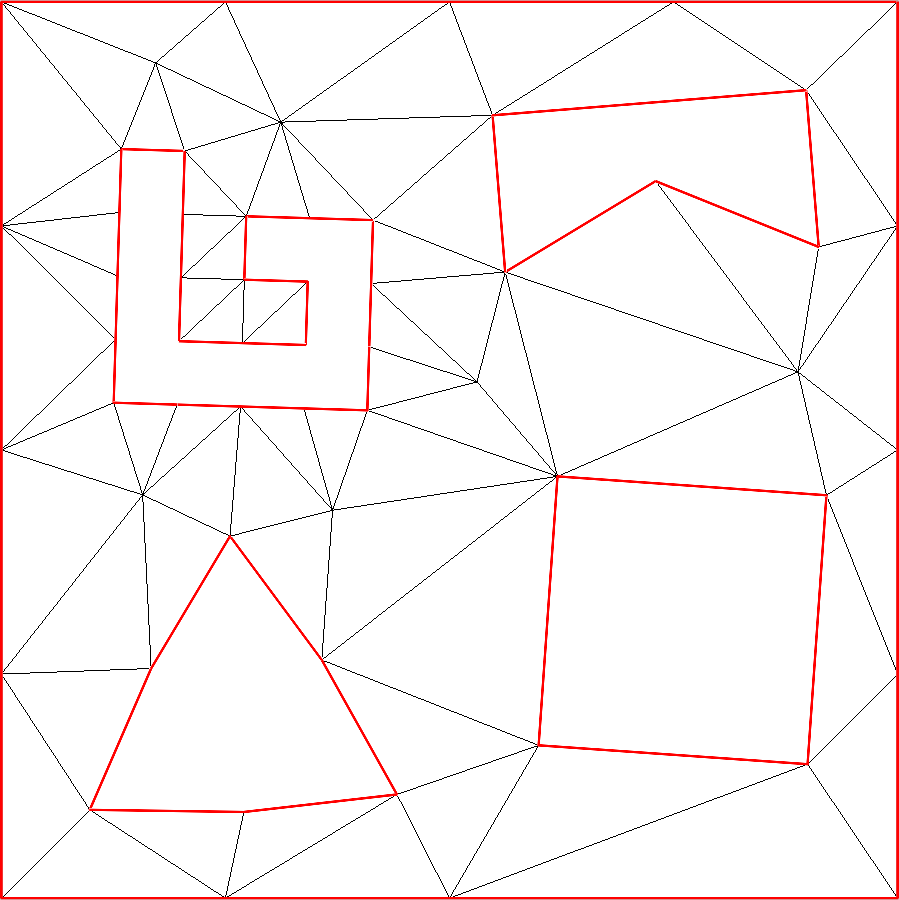}
&\includegraphics[width=.22\textwidth]{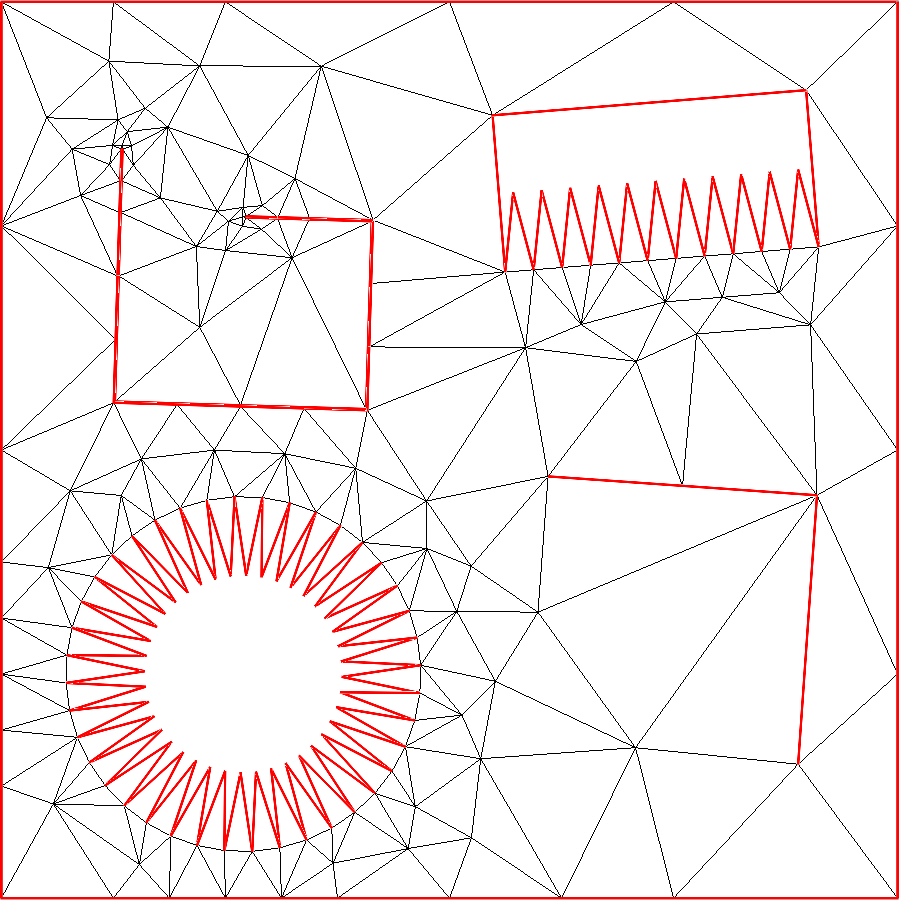}
\end{tabular}
\caption{Multi-parametric polygon meshes.\label{fig:multi-poly-parametric}}
\end{figure}

To generate new polygon meshes from scratch, the user is asked to select one or more polygons to be added to the domain. 
Two types of predefined polygons are made available for user selection: parametric and random~\cite{CHANGE}. When a set of random polygons is selected, the generated data set is made of a single mesh; if at least one parametric polygon is chosen, then a family of meshes \mbox{$D=\{M(0),\ldots,M(1)\}$}, is generated, where~$M(0)$ contains all the parametric polygons at its initial phase (e.g., they do not present critical geometric features (Fig.~\ref{fig:multi-poly-parametric}(a)) and they are progressively made worse by a deformation, controlled by the parameter~$t\in[0,1]$ (Fig.~\ref{fig:multi-poly-parametric}(b)). In the latter case, the number of generated meshes is user-defined.

Other than the predefined parametric polygons
, the user is allowed loading polygons from a file. Such a polygon is automatically scaled and translated to be placed inside the domain, and additional editing can be applied by the user. 
Users are allowed to save the polygon configuration into a CSV file to be possibly reloaded during any following experimental session. The CSV stores, for each polygon, any data necessary to rebuild the configuration (i.e., position, scale, rotation).
%
%
\subsection{PEM solver\label{sec:pem-requirements}}
As aforementioned, 
\frameworkName does not include PEM solvers but 
is intended to be a support for the analysis of any kind of PEM solvers, with no limitation. The only requirement to be satisfied is related to the solver output format, which must be available according to a very simple text file format. Specifically, both the numerical solution computed by the VEM solver and the ground-truth solution (if any) must be as a list of their values at the mesh vertices. Finally, these two arrays are saved in a \textit{.txt} file whose name is composed of the input filename and an additional ending to indicate which output it encodes (i.e. either \textit{``-solution''} or \textit{``-ground-truth''} respectively). For each performance evaluation, an additional \textit{.txt} file is generated and its name is composed of the input filename and an additional string to indicate the performance name. The single value representing the solver performance must be written in the file. 
%

\subsubsection{Example of an Application}
As a matter of example and test, we exploit the family of parametric and random polygon meshes defined in \cite{CHANGE} as an input dataset and the \emph{Virtual Element Method} (VEM)~\cite{basicVEM} as an external PEM solver. Specifically, we considered a test problem corresponding to solutions to the Poisson equation in the domain $\Omega = (0,1)^2$. For our test case, the ground truth is the Franke function, namely

\begin{equation}
\begin{split}
	u_2(x,y):=\frac 3 4 e^{-\left(
		(9x-2)^2+(9y-2)^2
		\right)/4}\\
		+ \frac 3 4 e^{-
		\left(
		(9x+1)^2/49 + (9y+1)/10\right)}\\
	+ \frac 1 2 e^{-
		\left(
		(9x-7)^2 + (9y-3)^2	
		\right)/4}\\
	+\frac 1 5 e^{-
		\left(
		(9x-4)^2 + (9y-7)^2
		\right)
	}
\end{split}
\end{equation}

The MATLAB\textsuperscript{\textregistered} code of the method computes the PDE solution, provides the solution ground truth and also computes some evaluations of PEM solver quality, such as 
\begin{itemize}
\item~$\mathcal{L}_{1}$ \emph{condition number} \mbox{$\kappa_{1}(\mathbf{S}) = \|\mathbf{S}\|_1\|\mathbf{S}^{-1}\|_1$} \emph{of the PEM stiffness matrix}~$\mathbf{S}$;
\item \emph{relative error} \mbox{$\epsilon_S:=\|\mathbf{u}-\mathbf{u_h}\|_S/\|\mathbf{u}\|_S$}, with weighted norm \mbox{$\|\mathbf v\|_S^{2} = \mathbf v^{\top}\!\mathbf{S}\:\mathbf v$};
\item \emph{relative~$\mathcal{L}_{\infty}$-error} \mbox{$\epsilon_{\infty}:=\|u-u_h\|_{\infty}/\|u\|_{\infty}$}, between the ground-truth~$u$ and the computed~$u_{h}$ solutions.
\end{itemize}

\begin{figure}[h]
\centering
\begin{tabular}{c}
(a)\includegraphics[width=.38\textwidth]{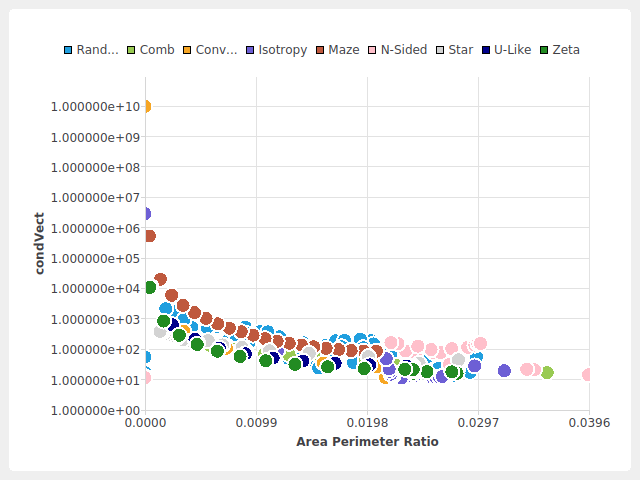}\\
(b)\includegraphics[width=.38\textwidth]{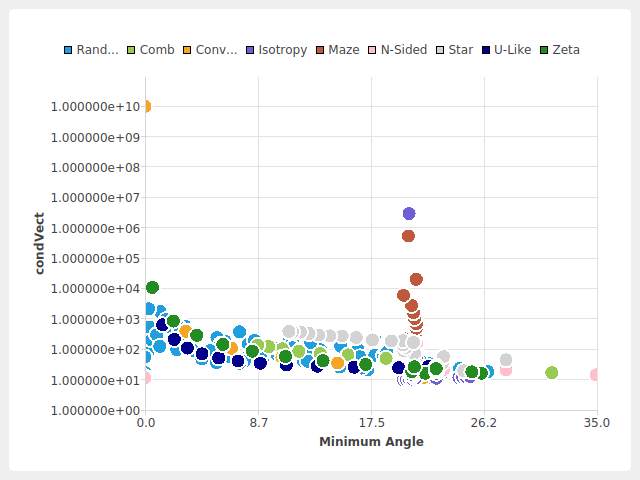}
\end{tabular}
\caption{Scatter plot of the correlation between geometric properties and performance metrics of the PEM solver: correlation between ($x$-axis) (a) the area-perimeter ratio and (b) the minimum angle with ($y$-axis) the conditioning number of the stiffness matrix of the PEM solver. \label{fig:scatterplot}}
\end{figure}

To enable the visualization of the results and the correlation between PEM solver performances and geometric properties, we wrap the code into a MATLAB function to be called from a command line and we redirect the output of the \emph{Virtual Element Method} to file, according to the file format described in Sec.~\ref{sec:pem-requirements}. The MATLAB function requires both the input mesh and the output directory as parameters. These two simple operations are sufficient to make our framework 
and the \emph{Virtual Element Method} communicate. Fig.~\ref{fig:solver_results} shows how \frameworkName visualizes PEM solver performances, while scatter plots in Fig.~\ref{fig:scatterplot} are generated by \frameworkName and show the correlation between two geometric metrics (area perimeter ratio and minimum angle) and the conditioning number of the stiffness matrix of the PEM solver.

%

\section{Discussion and conclusions\label{sec:conclusions}}
We presented a novel tool helping researchers to perform experimental design and analysis of polygonal meshes for PEM solvers. Its easy-to-use graphical interface simplifies the execution of experimental pipelines, from the design of polygon meshes stressing specific geometric properties to the analysis of performances of user-provided PEM solvers. It also allows correlating geometric properties with PEM solver performances and provides advanced visualization modalities of the results of such an analysis. \frameworkName 
is available as an open-source project (\url{https://github.com/DanielaCabiddu/PEMesh}) and we expect it can be employed in several research activities.
%
%
\paragraph*{Current limitations and future works}
There are several directions in which our framework can be improved. First, the current implementation allows the definition of polygon meshes by loading already exiting polygons, possibly designed by exploiting external tools. Since our proposal is intended to support activities in several research fields, additional features, such as the possibility to draw polygons freehand, would simplify the mesh generation process for users coming from fields other than mesh design. A deeper analysis, including user studies, can support the development of an improved version of the tool according to user needs.

Also, \frameworkName 
supports 2D meshes, but the entire architecture is agnostic to the dimension of the geometric input. It is almost trivial to extend the graphical user interface to support 3D meshes, but future investigations are necessary on 3D mesh generation approaches and the definition of geometric properties that would be likely to be of interest to the research community.

Finally, \frameworkName 
is designed as a desktop application exploiting RAM to both generate meshes and run PEM solvers. At each run of the application, the complete input dataset is kept in core to improve efficiency. Nevertheless, the power of \frameworkName is RAM-bounded and it is not robust enough to support arbitrary large input datasets (i.e. made up of a large number of large meshes). Future activities will be addressed to improve the underlying architecture and guarantee efficiency independently of the input dataset size.
%


\section*{Acknowledgments}

This document is the results of the research project funded by the ERC Project CHANGE, which has received funding from the European Research Council (ERC) under the European Union, Horizon 2020 research and innovation programme (grant agreement Nr. 694515).


\printbibliography                



\end{document}